\documentstyle[epsf]{mn}

\title[$UBV$ stellar photometry in M5. II. Physical parameters of HB stars]
 {$UBV$ stellar photometry of bright stars in GC M5. II. Physical
parameters of HB stars}
\author[P.V. Baev et al.]{P. V. Baev,$^{1}$ 
 H.~Markov,$^{2}$ and N. Spassova$^{3}$ \\
\\
$^{1}$Merrowinger str. 10, D-40223 D\"{u}sseldorf, Germany, e-mail: pvbaev@rambler.ru\\
$^{2}$Natl. Astr. Obs. `Rozhen', IA, BAS, p.o.box 136, BG-4700
Smoljan, Bulgaria, e-mail: rozhen@mbox.digsys.bg\\
$^{3}$Institute of Astronomy, Bul. Acad. of Sci., 72 Tsarigradsko
ch., BG-1784 Sofia, Bulgaria, e-mail: neda@astro.bas.bg\\
}
\date{Received 2000 June 15; in original form 1999 June30}
\begin{document}

\maketitle

\begin{abstract}
The physical parameters of the stars in the central region of the
globular cluster M5 (NGC~5904) were determined from $UBV$
photometry using Kurucz (1992) (K92) synthetic flux distributions
and some empirical relations. It is found that the bluest
horizontal branch (HB) stars have higher luminosities than
predicted by canonical zero-age horizontal branch (ZAHB) models.
Parameters of the mass distribution on the HB stars are
determined. It is shown that the gap in the blue HB previously
reported (Markov et al. 1999, hereafter Paper~I) is probably a
statistical fluctuation.
\end{abstract}

\begin{keywords}
globular clusters: general -- globular clusters: individual
(M5$\equiv$NGC5904) -- stars: evolution -- stars: horizontal
branch -- stars: Population~II.
\end{keywords}

\section{Introduction}

Crocker, Rood \& O'Connel (1988, CRO88) determined
spectroscopically $\log T_{\mathrm{eff}}$ and $\log g$ for $14$
BHB stars in globular cluster (GC) M5. They found that M5 BHB
stars matched the canonical ZAHB models quite well. In our
photometry (Paper~I) there are $11$ stars common with CRO88.
Assuming a relationship between the temperatures and surface
gravities as derived in CRO88 and the atmospheric models of K92,
we are able to obtain the physical parameters for more than $100$
BHB stars from our photometric observations.

According to Buonanno et al. (1981) (BCF81) and Crocker \& Rood
(1985) the bluest BHB stars tend to rise above the canonical ZAHB
whilst Bohlin et al. (1985) did not find such effect. The
completeness of our sample throws additional light on this matter.

The present work is structured as follows. In \S \ref{physical},
the transformation of observed colours and magnitudes to the
($\log T_{\mathrm{eff}}$,$~\log L$) plane is described. In \S
\ref{synthetic}, an approximation of the optical colour--magnitude
diagram (CMD) in terms of synthetic HBs is presented and the
statistical importance of the gap in the blue HB discussed in
Paper~I is evaluated. \S \ref{BHB_stars} is dedicated on the
position of BHB stars in the theoretical diagram and finally our
findings are summarized in \S \ref{summary}.

\section{Physical characteristics of HB stars}

We determine the effective temperatures of BHB stars from their
reddening-free colours. For this purpose, we use theoretical
models of energy distributions for different temperatures,
gravities and abundances, in particular the line blanket
atmospheres of K92. We choose the model with metallicity
[Fe/H]$=-1.5$, i.e. the nearest value to that derived in Paper~I
($-1.38$). It is well known that the model grids make folds on the
$(U-B)_{0}$,~$(B-V)_{0}$ and $(U-V)_{0}$,~$(B-V)_{0}$ planes. This
leads to a nonsingle-valued transformation from colours to
temperatures and gravities. Unfortunately, all the BHB stars of M5
are situated just within the folds and we forfeit the opportunity
for simultaneous determination of $\log T_{\mathrm{eff}}$ and
$\log g$. After a suitable gravity-averaging of the atmospheric
parameters, it becomes possible to achieve the necessary
unequivocal dependence of the colours on the effective
temperature.

The analysis of the data for BHB stars is divided into two parts.
We consider separately hot BHB stars (those with
$(B-V)_{0}<-0.02$) and cool BHB stars (with $(B-V)_{0}>-0.02$).
The reason is that the different colour indices as temperature and
gravity indicators exchange their role at about
$T_{\mathrm{eff}}=9\,200\mathrm{K}$. The $(U-B)_{0}$ colours as a
measure of the Balmer discontinuity and $(U-V)_{0}$ as a long
baseline colours are very sensitive temperature parameters for
B--type stars in the $UBV$ system. The $(B-V)_{0}$ colour is a
good temperature indicator for cooler stars ($\log
T_{\mathrm{eff}}<3.96$). For that reason different models are
adopted to analyse the hot and cool BHB stars. \label{physical}

\subsection{Determination of $\log T_{\mathrm{eff}}$ for hot BHB
stars}

The effective temperature is derived from the observed $UBV$ data
using the relations $Q(\log T_{\mathrm{eff}})$ and $(U-V)_{0}(\log
T_{\mathrm{eff}})$ found in K92. These relations are applied for
$Q<0.0$ and $(U-V)_{0}<0.10$. Both relations are employed for
$\log g$ in the interval $(3,\,4.5)$ and then the derived
$T_{\mathrm{eff}}$ values are averaged. The effective temperatures
for $67$ BHB stars are determined in this way. The dispersion in
$\log T_{\mathrm{eff}}$ in the two cases is $\pm 0.013$ and $\pm
0.020$. The mean standard errors of $Q$ and $(U-V)_{0}$ are about
$\pm 0.04$ mag, so the error in $\log T_{\mathrm{eff}}$ should not
exceed $\pm 0.043$ ($\pm 1\,100$\textrm{K} at
$T_{\mathrm{eff}}=10\,000$\textrm{K}). \label{sect21}

\subsection{Determination of $\log T_{\mathrm{eff}}$ for cool BHB
stars}

$Q$ and $(U-V)_{0}$ are no longer sensitive temperature indicators
for stars in this evolutionary stage. $(B-V)_{0}$ with a maximum
gravity dependence of $\Delta (B-V)_{0}=0.18$ at $\log
T_{\mathrm{eff}}=3.9$ for $\log g$ in the interval $(2,\,4.5)$ is
a more appropriate indicator. An assumption about $\log g$ is
needed to derive the temperatures. Model computations show that
$(U-B)_{0}$ is a strong gravity indicator for these stars. This is
demonstrated in Fig.~7 of Paper~I, where the theoretically
predicted two-colour relations taken from K92 are overlaid on the
observed data. As we discussed in \S 4 of Paper~I, the majority of
these HB stars show an ultraviolet deficiency and their positions
are consistent with the loci of the models with ${\log g\sim
2.0}$. For this reason $\log g$ within the interval $(2,\,3.5)$ is
chosen here. The model suggests a mean standard deviation for
$\log T_{\mathrm{eff}}$ of $\pm 0.021$. Taking into account that
the uncertainty in $(B-V)_{0}$ is about $\pm 0.04$ mag, we find
the error in $\log T_{\mathrm{eff}}\sim \pm 0.045$, which
corresponds to $900\mathrm{K}$ at
$T_{\mathrm{eff}}{=}8\,000\mathrm{K}$. \label{sect22}

\begin{figure}
\epsfysize=4.3cm \epsfbox{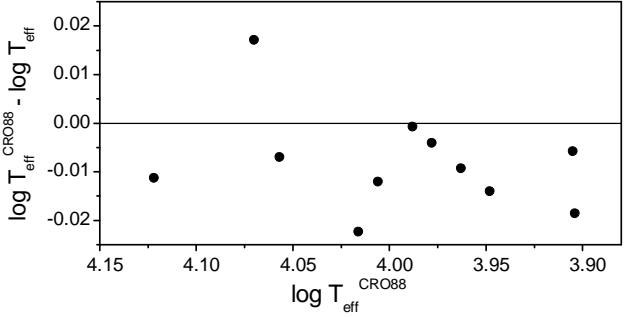}%
%
\caption{Comparison between the effective temperatures $\log
T_{\mathrm{eff}} $ derived in this work and the CRO88 values for
$11$ common stars.} \label{Fig1_Cmprsn}
\end{figure}

The effective temperatures of $111$ hot and cool HB stars are
obtained using this technique. A check of the validity of the
determined temperatures is made by comparison with $11$ BHB stars
in common between the CRO88 and our observations.
Figure~\ref{Fig1_Cmprsn} illustrates a tendency of our
temperatures to be higher on average by $0.008\pm 0.003$.

\begin{figure}
\epsfysize=4.4cm \epsfbox{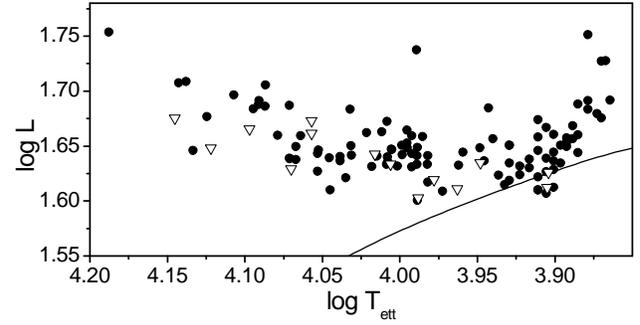}%
%
\caption{$\left( \log L,\,\log T_{\mathrm{eff}}\right) $ diagram for all $%
111 $ BHB stars (solid circles) with $UBV$ magnitudes in our
sample. The open \ triangles denote $14$ stars of CRO88. The solid
line is the ZAHB of Dorman et al. (1993).} \label{Fig2_logL_logT}
\end{figure}

Provided that the effective temperatures are determined,
appropriate K92 relations ($BC\left( \log T_{\mathrm{eff}}\right)
$ and $BC\left[ (B-V)_{0}\right] $ for hot and cool BHB stars,
respectively) are used to derive bolometric corrections (BC), and
then bolometric magnitudes and luminosities. The derived
luminosities are based on the assumption that
$(m-M)_{V_{0}}=14.46\pm 0.03$ and $M_{\mathrm{bol}}^{\sun }=4.72$.
The best fit of our observed CMD to the canonical ZAHB of Dorman
et al. (1993) (DRO93) is used to obtain the distance modulus. The
DRO93 evolutionary models are for [Fe/H]$=-1.48$, [O/Fe]$=0.63$,
$Y_{HB}=0.25$ and core mass $M_{\mathrm{core}}=0.485$\thinspace
$M_{\sun }$ (hereafter we imply only this set of DRO93 tracks when
we refer to the evolutionary theory). The mean square error of the
luminosity determinations is $\pm 0.1$ in $\log L$. The ($\log
L$,\thinspace $\log T_{\mathrm{eff}}$) diagram for all $111$ BHB
stars with $UBV$ magnitudes in our sample is given in Fig.~\ref
{Fig2_logL_logT}.
The stars of CRO88 with spectroscopically measured atmospheric
characteristics are also plotted as open triangles.

\subsection{Determination of the stellar masses}

Knowing the effective temperatures, luminosities and surface
gravities of the stars, we can estimate the corresponding masses
independently of their evolutionary history through the relation
$M/L\sim g/T_{\mathrm{eff}}^{4}$ or in more convenient form
\[
\log \left( M/M_{\sun }\right) =\log \left( L/L_{\sun }\right)
-4\log T_{\mathrm{eff}}+\log g+10.611.
\]

We rely on a sample of $12$ BHB stars with reliable $\log g$ and
$\log T_{\mathrm{eff}}$ data (CRO88) to derive a linear relation
between surface gravity and effective temperature. A least square
fit yields:

\begin{equation}
\log g=-11.244(\pm 1.929)+3.699(\pm 0.478)\,\log
T_{\mathrm{eff}}\,. \label{eq1}
\end{equation}
Equation (\ref{eq1}) is deduced for stars with $\log
T_{\mathrm{eff}}$ in the interval $(3.948,\,4.145)$, and it is
used to obtain the surface gravities of all BHB stars with r.m.s.
$\sim 0.15$, which corresponds to the errors in the input data.

\begin{figure}
\epsfysize=4.8cm \epsfbox{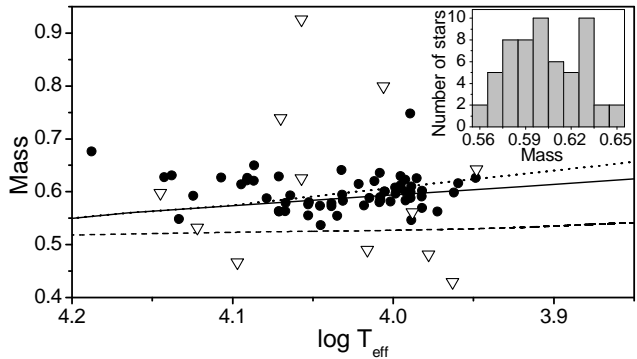}%
%
\caption{(mass,$\,\log T_{\mathrm{eff}}$) diagram. The solid
circles are stars with obtained masses. The $12$ stars of CRO88
are plotted as open triangles. The solid line is the (ZA
mass,$~\log T_{\mathrm{eff}}$) relation; the dotted line is the
relation between the stellar mass and maximal temperature
reachable during the post-ZAHB evolution; the dashed line is the
isochrone at $t=110\,\mathrm{Myr}$. The inset shows the mass
distribution.} \label{Fig3_mass_logT}
\end{figure}

The errors of the mass determination consist of both the errors of
$\log T_{\mathrm{eff}}$ and $\log g$, and those of $\log L$.
Assuming errors in $\log T_{\mathrm{eff}}$ of $\pm 0.043$, in
$\log g$ of $\pm 0.15$, and in $\log L$ of $\pm 0.1$ we obtain a
mean error in the mass derivation of approximately $\pm 0.19$.
Changes in the adopted distance modulus and extinction would
result in a systematic shift of the derived masses. The dependence
of the stellar mass on the effective temperature is shown in
Fig.~\ref {Fig3_mass_logT}. The distribution of the masses of $62$
stars situated within the temperature range ($8\,800\,$\textrm{K},
$14\,000\,$\textrm{K}) of validity of the linear relation $\log
g\left( \log T_{\mathrm{eff}}\right) $ is plotted in an inset on
Fig.~\ref{Fig3_mass_logT}.

During their HB lifetime, the stars reach a maximal effective
temperature (see, for instance, DRO93). This temperature as a
function of stellar mass is also represented in
Fig.~\ref{Fig3_mass_logT} by dotted line. It almost completely
coincides with the isochrone at\thinspace\ $t=60\,$\textrm{Myr}.
Although the error of the stellar masses is considerable,
Figure~\ref {Fig3_mass_logT} implies that the masses of the
hottest stars ($\log T_{\mathrm{eff}}>4.07$) are systematically
higher than those corresponding to the limit temperatures. This
suggests a discrepancy in luminosity compared with theory. The
masses of the $12$ stars of CRO88 are also plotted in
Figure~\ref{Fig3_mass_logT} (open triangles). The big scatter of
these stars around the zero-age relation can be interpreted as an
inaccurate determination of the surface gravities (see Table~2 of
CRO88).

\section{HB synthetic ($M_{V}$,~$(U-V)_{0}$) diagram}

\label{synthetic}

\subsection{Parameterisation of the HB}

A curve (ridge line), which is close enough to a HB strip, is
determined by means of averaging the stellar positions on small
segments covering the HB. A position along the HB, from the red
extreme toward the blue end, is given by arc length ($\ell
_{\mathrm{HB}}$) of the normal projection on the ridge line (for
similar procedures, see Rood and Crocker 1989; Ferraro et al.
1992; Dixon et al. 1996).

\begin{figure}
\epsfysize=4.5cm \epsfbox{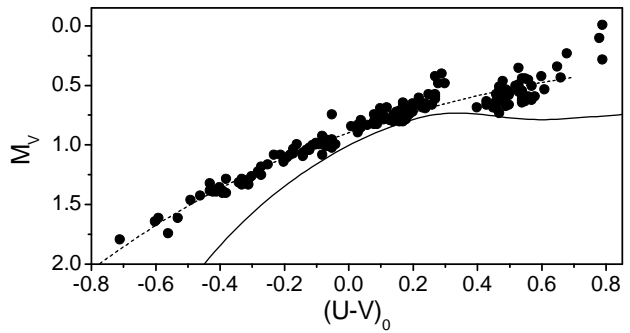}%
%
\caption{Colour--magnitude diagram of the HB region of the M5 and
the ridge line (dashed line) used to build the stellar
distribution along the HB. Solid line is the ZAHB of DRO93.}
\label{Fig4_CMD}
\end{figure}

\begin{figure}
\epsfysize=8.0cm \epsfbox{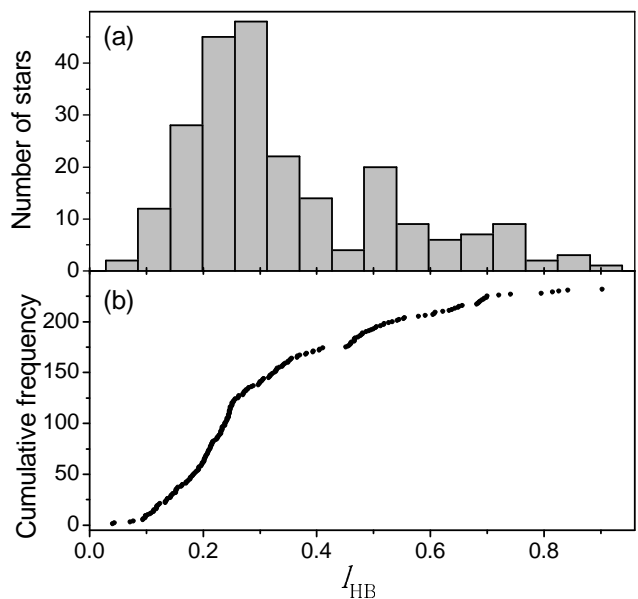}%
%
\caption{The observed stellar distribution along the HB in M5. The
histogram of the $\ell _{\mathrm{HB}}$--coordinate of the stars is
represented on panel (a) and the cumulative frequency -- on panel
(b).} \label{Fig5_l_distr}
\end{figure}

The ridge line and HB region of M5 is shown in Fig.~\ref{Fig4_CMD}
(dashed line).
To make possible a comparison of our distribution to other ones,
we give the scale ratios $\Delta \ell _{\mathrm{HB}}/\Delta
(U-V)_{0}=3.341$ and $\Delta \ell _{\mathrm{HB}}/\Delta
M_{V}=2.468$. The values of the HB red extreme of the ridge line
are $M_{V}^{\mathrm{RE}}=0.43$ and\thinspace\
$(U-V)_{0}^{\mathrm{RE}}=0.7$, the blue and red edges of the
instability strip at the level of the ridge line are
$(U-V)_{0}^{\mathrm{IS}_{\mathrm{B}}}=0.26$ and\thinspace\
$(U-V)_{0}^{\mathrm{IS}_{\mathrm{R}}}=0.45$ (see Fig.~\ref
{Fig4_CMD}). The observed distribution is shown in
Fig.~\ref{Fig5_l_distr} both as a histogram and a cumulative
frequency.
The $\ell _{\mathrm{HB}}$--values of the blue and red edges of the
instability strip are $\ell
_{\mathrm{HB}}^{\mathrm{IS}_{\mathrm{B}}}=0.244$ and $\ell
_{\mathrm{HB}}^{\mathrm{IS}_{\mathrm{R}}}=0.143$. \label{sect31}

\subsection{Theoretical models of the HB}

To create a model HB, we transform the mass-age relation (after
ZAHB) into a colour-magnitude plane using the theoretical
correspondence between the (mass, age) and ($\log L$,$~\log
T_{\mathrm{eff}}$) or ($\log g$,$~\log T_{\mathrm{eff}}$)
relations. Evolutionary tracks (DRO93), synthetic atmospheres
(K92), the adopted metallicity, and distance modulus are the same
as in \S \ref{physical}. A Hermite interpolation (Kahaner et al.
1988) of the DRO93 tracks is employed. We adopt both a unimodal
(Gaussian) mass distribution and a bimodal one (see e.g. Rood
1973; Lee et al. 1990; Dixon et al. 1996).

An assembly of $76$ RRLyrae stars randomly distributed in the
instability strip in ($M_{V}$,~$(B-V)_{0}$) plane with a mean
magnitude $\langle V_{\mathrm{RR}}\rangle =15.09$ (Reid 1996) is
added. From the CMD, the blue edge of the instability strip is
located at $(B-V)_{0}^{\mathrm{IS}_{\mathrm{B}}}=0.16$ and the red
edge -- at $(B-V)_{0}^{\mathrm{IS}_{\mathrm{R}}}=0.38$ (for
comparison see BCF81). The adopted number of variable stars is
derived under condition that the sample represents $74$ per cent
of all HB stars located in the measured region.

Taking into account the weak dependence of the HB lifetime on the
total mass of the stars, we can introduce a universal HB lifetime
$\sim 110$\thinspace \textrm{Myr}. Due to the assumption of a
constant rate of populating the ZAHB, the ages after the ZAHB are
considered as random numbers uniformly distributed in the range
$0$--$110$\thinspace \textrm{Myr}. We also presume known stellar
mass distribution as well as age distribution. For a unimodal mass
distribution (UMD) the transformation method (Box--Muller
algorithm) is used to produce the Gaussian deviates, whereas in
the case of a bimodal mass distribution (BMD) the appropriate
deviates are generated by means of the rejection method. Normally
distributed random `observational' errors are introduced at the
end of the mapping.

To restrict the unknown parameters of the mass distribution that
produce the observed CMD, we compare statistically the
corresponding distributions of synthetic and observed HBs. A
two-sample Kolmogorov--Smirnov (KS) test is
used to check the null hypothesis that model and observed cumulative $\ell _{%
\mathrm{HB}}$--frequencies ($\ell _{\mathrm{HB}}$--curves) are
drawn from the same parent distribution. Simultaneously a
chi-square ($\chi ^{2}$) test is applied to binned $\ell
_{\mathrm{HB}}$--coordinate distributions. Each synthetic HB
contains the same number of stars as our observed sample. At the
$99$ per cent level of significance, we disprove the null
hypothesis for all parameters outside the regions listed in
Table~\ref{Table1}.
\begin{table}
\begin{tabular}{llll}
\hline Model & Parameter & Permissible interval & Best fit \\
\hline\hline UMD & $\left\langle M_{\mathrm{HB}}\right\rangle $ &
$\left(
0.5801,0.6782\right) M_{\sun }$ & $0.6327M_{\sun }$ \\
& $\sigma _{M_{\mathrm{HB}}}$ & $\left( 0.0179,0.0433\right) M_{\sun }$ & $%
0.0321M_{\sun }$ \\ \hline BMD & $\left\langle
M_{\mathrm{HB}}\right\rangle _{1}$ & $\left(
0.5699,0.6432\right) M_{\sun }$ & $0.6057M_{\sun }$ \\
& $\sigma _{\left( M_{\mathrm{HB}}\right) ,1}$ & $\left(
0.0181,0.0199\right) M_{\sun }$ & $0.0192M_{\sun }$ \\
& $\left\langle M_{\mathrm{HB}}\right\rangle _{2}$ & $\left(
0.6110,0.6850\right) M_{\sun }$ & $0.6493M_{\sun }$ \\
& $\sigma _{\left( M_{\mathrm{HB}}\right) ,2}$ & $\left(
0.0258,0.0288\right) M_{\sun }$ & $0.0274M_{\sun }$ \\
& $f_{1}$ & $\left( 0.3703,0.4054\right) $ & $0.3898$ \\ \hline
\end{tabular}
\caption{Permissible intervals of the parameters determining the
model mass distributions and their best-fitting values.}
\label{Table1}
\end{table}
$\left\langle M_{\mathrm{HB}}\right\rangle $ and $\sigma
_{M_{\mathrm{HB}}}$ are the mean value and the standard deviation
of the Gaussian mass distributions, respectively. In the case of
bimodal mass distribution, we have an additional parameter
$f_{1}$, the fraction of HB stars contributed by the first
Gaussian. The mean values of the significant levels $\left\langle
Q_{\mathrm{KS}}\right\rangle $ and $\left\langle Q_{\chi
^{2}}\right\rangle $ are obtained by generating $\sim 250\,000$
synthetic HBs with the best-fitting parameters (see
Table~\ref{Table1}). When the UMD is used $\left\langle
Q_{\mathrm{KS}}\right\rangle =0.43$ and $\left\langle Q_{\chi
^{2}}\right\rangle =0.17$, whereas in the case of BMD
$\left\langle Q_{\mathrm{KS}}\right\rangle =0.48$ and
$\left\langle Q_{\chi ^{2}}\right\rangle =0.25$, which makes the
second distribution preferable.

\begin{figure}
\epsfysize=11.95cm \epsfbox{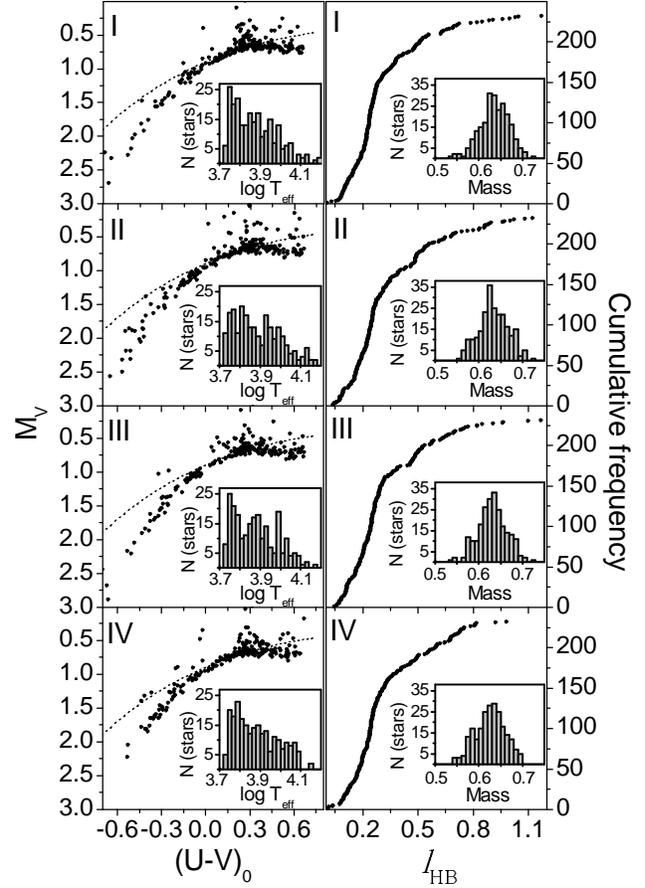}%
%
\caption{Four synthetic HBs randomly chosen from the set
of\thinspace\ $\sim 250\,000$\thinspace\ HB simulations and
cumulative frequency of $\ell _{\mathrm{HB}}$--coordinate, which
are computed by unimodal mass distribution. The insets show the
histograms of the $T_{\mathrm{eff}}$ and mass distributions. The
dashed line is the ridge line.} \label{Fig6_Sy_UMD_CMD}
\end{figure}

\begin{figure}
\epsfysize=11.95cm \epsfbox{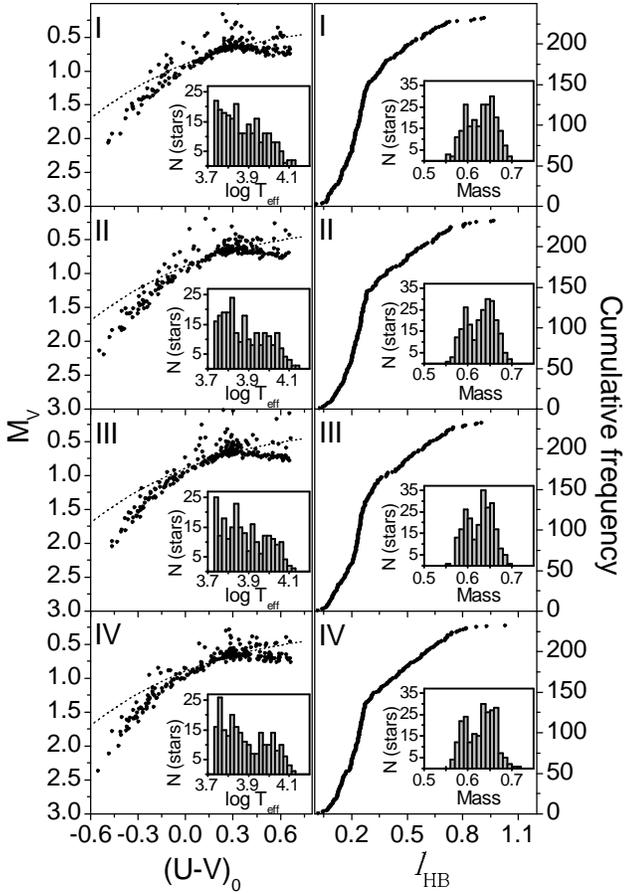}%
%
\caption{Same as in Figure~\ref{Fig6_Sy_UMD_CMD}, but for the
bimodal mass distribution.} \label{Fig7_Sy_BMD_CMD}
\end{figure}

Figure~\ref{Fig6_Sy_UMD_CMD} shows four synthetic HBs computed by
means of UMD, as well as the corresponding cumulative
distributions of coordinates, the histograms of masses and
temperatures.
Similar results are plotted on Figure~\ref{Fig7_Sy_BMD_CMD} for
bimodal mass distribution.
In both cases (UMD and BMD) the corresponding four synthetic HBs
are chosen randomly from the whole sets of\thinspace \thinspace\
HB simulations. At first glance, there are considerable
differences between the separate synthetic HBs although the model
assumptions are just the same. The principal cause is that the
random deviates used for the mass generation only asymptotically
approach the desired distributions. However, a thorough inspection
reveals that the blue tails of UMD synthetic HBs are less
populated and more extended than tails of the BMD synthetic HBs --
a prospective effect due to the bigger distribution tail area of
the unimodal mass distribution.

The main discrepancy between the observed and model HBs is the
downward deviation of the synthetic blue tails from the ridge
line. Both UMD and BMD computations show the same tendency. The
formal cause may be the inconsistency of the uniform distribution
of the stellar ages after ZAHB adopted in the computations,
although, most probably, the real cause is the unreliability of
canonical evolution theory. \label{sect32}

\subsection{Gaps on the blue HB}

Even a passing glance at the simulated HBs ascertains substantial
statistical fluctuations, which modify their morphology. The
observed gap discussed in Paper~I may thus be such a fluctuation.
We consider as gaps regions with a local minimum of the slope of a
monotonically increasing cumulative frequency function. The
standard procedure to test the statistical significance of
suspected gaps is based on the assumption that the probability to
find stars along the cumulative distribution is uniform with a
value $P_{L}$ left of the gap, and uniform with another value
$P_{R}$ right of the gap (Aisenman et al. 1969). Usually, the
significance of this hypothesis is not taken into account.
Instead, one verifies whether the probability of finding stars in
the gap substantially differs from the mean of the probabilities
$P_{L}$ and $P_{R}$ (Howarden 1971). When applied to cumulative
$\ell _{\mathrm{HB}}$-distributions of HB stars, the standard
method yields a local highly overestimated statistical
significance of the gaps. In the present work, we apply KS-test to
a section of $\ell _{\mathrm{HB}}$-curve in length of five gap
widths in order to examine the hypothesis of uniform distribution
around and in the gap (probably with three different densities).

\begin{figure}
\epsfysize=13.0cm \epsfbox{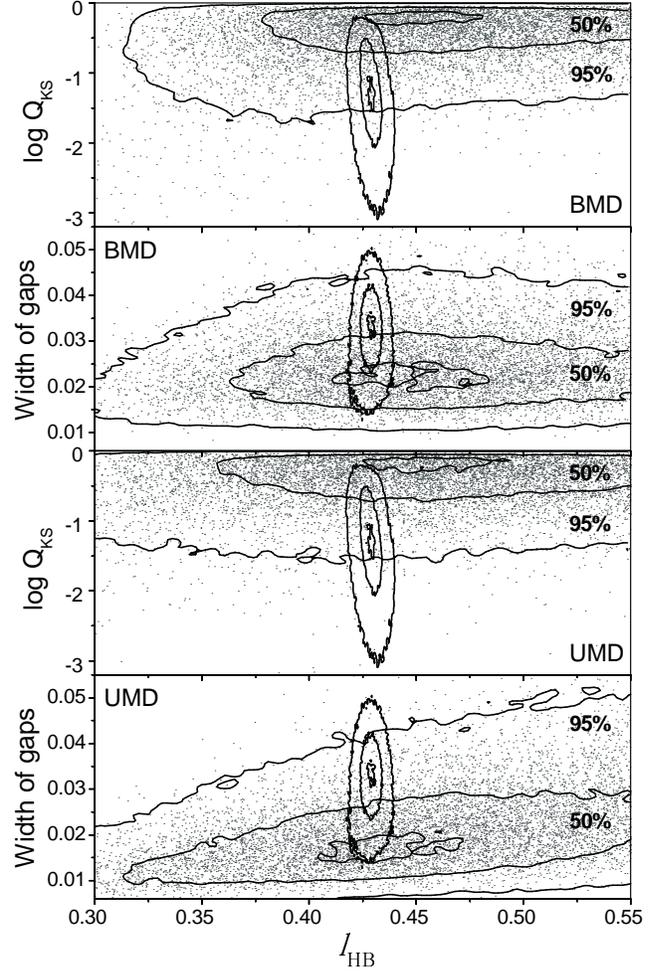}%
%
\caption{Characteristics of the observed and detected gaps on the
synthetic HBs computed for UMD and BMD.} \label{Fig8_Gap_stat}
\end{figure}

The observed gap should be represented as a domain on the ($W$,
$\ell _{\mathrm{HB}}$) plane ($W$ is the gap width) because of
photometric errors. To derive the borders of this region, we vary
the stellar magnitudes and colours within photometric accuracy
bounds reported in Paper I, and produce a set of $\sim 200\,000$
modified observed HBs. Then, we obtain the joint probability
density function (PDF) of the observed gap $f_{o}\left( W,\ell
_{\mathrm{HB}}\right)$ using an automatic search and inspection of
arising intervals. Three elliptic-like contours (with major axes
in ordinate direction) containing $95$, $50$, and $5$ per cent of
all detected gaps are plotted on Figure~\ref{Fig8_Gap_stat} to
depict this function.

As mentioned in \S \ref{sect32}, the difference between the models
and observed data is statistically non-significant. This enables
us to use the estimation of a mere chance probability that a gap
comparable with the observed one may appear on the synthetic HBs.
We consider the width and the location of the biggest gap on the
simulated $\ell _{\mathrm{HB}}$-curves as random variables. The
significant level $Q_{KS}$ of the gap region can also be
interpreted as a random parameter, which is a function of the gap
sharpness. We construct the sampling distributions empirically by
using the
set of $\sim 250\,000$ simulated HBs. The joint PDFs $f_{s}\left( W,\ell _{%
\mathrm{HB}}\right) $ and $f_{s}\left( Q_{KS},\ell
_{\mathrm{HB}}\right) $ are depicted on Figure~\ref{Fig8_Gap_stat}
as isodensity contours containing $95$, $50$, and $5$ per cent of
all revealed gaps. The gaps are also plotted as points.

To test the null hypothesis that there is no genuine difference
(within the limits of random variability) between the parameters
of observed and model gaps, we use the following three
probabilities (see Appendix~\ref{AppA}). $P_{1}$ and $P_{2}$ are
the averaged (over $f_{o}$) probabilities of arising of synthetic
gaps, which are wider than the observed one. They are calculated
by the marginal PDF ($m_{s}(W)$ for the ($W$, $\ell _{\mathrm{HB}}
$) relation and $m_{s}(Q_{KS})$ for the ($Q_{KS}$, $\ell
_{\mathrm{HB}}$) relation) and by a conditional PDF ($f_{s}(W|\ell
_{\mathrm{HB}})$ and $f_{s}(Q_{KS}|\ell _{\mathrm{HB}})$),
correspondingly. $P_{2}$ takes account of location on the $\ell
_{\mathrm{HB}}$-curve as distinct from $P_{1}$. $P_{3}$ is the
probability of appearance of synthetic gaps outside the region
enclosed in the averaged (over $f_{o}$) isodensity curve on which
the observed gap is situated. The so-defined probabilities for the
($W$, $\ell _{\mathrm{HB}}$) and ($Q_{KS}$, $\ell _{\mathrm{HB}}$)
relations are given in Table~\ref{Table2}.
\begin{table}
\begin{tabular}{rcccccc}
\hline
& \multicolumn{3}{c}{UMD model} & \multicolumn{3}{c}{BMD model} \\
Relation & $P_{1}$ & $P_{2}$ & $P_{3}$ & $P_{1}$ & $P_{2}$ & $P_{3}$ \\
\hline\hline
($W$, $\ell _{HB}$) & 0.21 & 0.10 & 0.30 & 0.20 & 0.20 & 0.48 \\
(log $Q_{KS}$, $\ell _{HB}$) & 0.08 & 0.08 & 0.16 & 0.09 & 0.08 & 0.17 \\
\hline
\end{tabular}
\caption{Probabilities used to test statistical significance of
the observed gap.} \label{Table2}
\end{table}
By the conventional standard of statistical significance
(probability $<0.05$), the observed gap is non-significant.
Therefore, the observed gap is not so extraordinary, as it seems
to be, and we tend to consider it as a statistical fluctuation.
\label{sect33}

\section{BHB stars and the theoretical H--R diagram}

In the ($\log L,~\log T_{\mathrm{eff}}$) plane
(Fig.~\ref{Fig2_logL_logT}), the tendency for the blue end of the
HB to rise above the ZAHB (the oxygen-enhanced models (DRO93) for
[Fe/H]$=-1.48$) is obvious. BCF81 and Crocker \& Rood (1985) have
found the same effect, using different methods and models for
transforming the observational data from a ($V$,$~\left(
B-V\right) $) plane into the theoretical ($\log L$,$~\log
T_{\mathrm{eff}}$) plane. On the other hand, Bohlin et al. (1985)
have not confirmed this trend (their Figure~7) when perform
ultraviolet observations with a rocket-borne telescope near
$1540$\AA \thinspace\ and \thinspace $2360$\AA \thinspace\ for
$50$ BHB stars.

An important argument for the validity of an upward trend is the
location in the ($\log L$,$~\log T_{\mathrm{eff}}$) diagram of
those $14$ BHB stars with atmospheric parameters being already
measured spectroscopically (CRO88). They are plotted in
Fig.~\ref{Fig2_logL_logT} as open triangles. On canonical theory
we expect the observed HB to be close to the ZAHB (DRO93). One
possibility for the revealed disagreement is that we have
underestimations of the adopted photometric errors. To check this
we use $11$ stars in common with CRO88. From the K92
temperature--colours calibrations we compute the corresponding
$(U-V)_{0}$, $(U-B)_{0}$ and $(B-V)_{0}$ values for each star. The
mean differences between the model and the observed colours are as
follows: $\Delta (U-V)_{0}=0.02$,\thinspace\ $\Delta
(U-B)_{0}=0.01$, and $\Delta (B-V)_{0}=0.02$ -- values comparable
within $1\sigma $ of our photometric errors.

\begin{figure}
\epsfysize=9.7cm \epsfbox{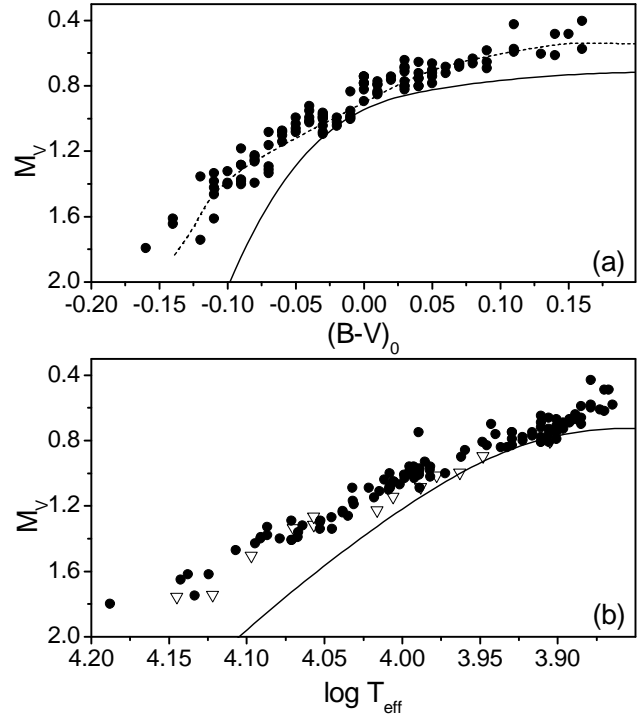}%
%
\caption{($M_{V}$,$~(B-V)_{0}$) diagram (panel~(a)) and
($M_{V}$,$~\log T_{\mathrm{eff}}$) diagram (panel~(b)) for BHB
stars with $UBV$ photometry. Open triangles denote the stars of
CRO88. Solid line is the ZAHB (DRO93). Dashed line on the
panel~(a) is the fiducial line of Sandquist et al. (1996). }
\label{Fig9_V_BV_V_logT}
\end{figure}

It can be seen from Fig.~\ref{Fig9_V_BV_V_logT}, where the
theoretical ZAHB is fitted to our observed data, that the blue end
of the observed HB (dots) do not match the model of DRO93 (solid
line).
This discrepancy has already been noted by Sandquist et al.
(1996). The fiducial line of BHB derived by these authors is
plotted for comparison as dashed line in
Fig.~\ref{Fig9_V_BV_V_logT}a. Sandquist et al. (1996) assume that
the disagreement results from the colour calibration. However, the
fact, that the high temperature stars with spectroscopically
obtained temperatures also have much higher luminosities (see
Fig.~\ref {Fig2_logL_logT}) than the predicted ones, means that
the observed effect cannot be explained by the colour calibration
only. In Fig.~\ref {Fig9_V_BV_V_logT}b, we compare the positions
of these stars (open triangles) (the $V$ mag for three of them are
taken from BCF81) with the location of the model HB. It is evident
that stars with $\log T_{\mathrm{eff}}>4.04$ deviate upward from
the model with $\Delta M_{V}$ in the range ($-0.2 $ to $-0.5$).
These values are too large to be explained with photometric
errors. A similar effect of `over--luminosity', reaching $\sim
0.4$\thinspace mag has been found by Grundahl et al. (1998) in the
globular cluster M13 using Stromgren photometry. Moreover,
according to Grundahl et al. (1999) all stars with
$T_{\mathrm{eff}}>11\,500\,$\textrm{K} in $14$ GCs demonstrate a
systematical deviation from the canonical ZAHB in the sense of
being brighter or hotter than the models. This effect is
independent of the metallicity, the mixing history on the RGB, and
other cluster parameters. The results in Fig.~\ref{Fig2_logL_logT}
are in a remarkable agreement with their results.

In explanation of this effect, we note that apart from the most
popular picture, in which HB stars are distributed along ZAHB
because of differences in the amount of mass loss during the
`He--flash' at the RGB tip, Fusi Pecci \& Renzini (1978) focused
attention on the possible effect of stellar rotation. With a
suitable choice of parameters the blue end of the `rotational'
ZAHB is expected to be more luminous than the non-rotating
canonical one.

The standard method used in \S \ref{physical} to obtain physical
parameters of the BHB stars does not permit simultaneous
determination of the effective temperature and the surface
gravity. The latter parameter is lost during the inevitable
averaging inherent to the elimination of the ambiguity of the
Kurucz grids (see \S \ref{physical}). For that reason we cannot
confirm independently the effect found in $6$ GCs including M5
(Moehler et al., 1995; Moehler, 1999), which consists in a
displacement from the ZAHB to lower $\log g$ of the bluest HB
stars having spectroscopically measured surface gravities. We
shall only remark that the linear relation (\ref{eq1}) shows the
same tendency to `under--gravity' towards the bluest HB stars in
M5. Grundahl et al. (1999) reveal the same connection between the
`over--luminosity' and the `under--gravity' in $14$ GCs, and
suppose that these effects are different manifestations of a
uniform physical mechanism. Relying on the spectroscopic data for
field and GC BHB stars, the authors (Grundahl et al. 1999)
hypothesize that this physical mechanism consists in radiative
levitation of elements heavier than carbon and nitrogen into the
stellar atmosphere, rather than a stellar interior or evolution
effects.

Formally, the possibility for a displacement toward higher
luminosities of some stars due to evolution off the ZAHB cannot be
ignored. Moreover, when the BHB stars reach the HB lifetime they
demonstrate the `under--gravity' effect simultaneously with the
`over--luminosity'. However, according to the canonical
evolutionary theory (DRO93) and the adopted concept of constant
evolution rate onto the ZAHB, there is no well-grounded reason
that so many low mass stars being at the same time in a transient
evolutionary phase as core helium exhaustion. \label{BHB_stars}

\section{Summary}

In the present paper we have applied standard methods to determine
the stellar physical parameters using a new $UBV$ stellar
photometry for the HB stars of the GC M5. Comparing the observed
$UBV$ colours with K92 synthetic atmospheres, we have derived the
effective temperatures and luminosities of a large sample of BHB
stars. The surface gravities have been derived assuming a linear
relation between spectroscopically measured $\log g$ and
$T_{\mathrm{eff}}$ of $12$ stars (CRO88).

We have found considerably high luminosities for the bluest HB
stars than predicted by the canonical theory.

Approximations of the distribution of stars in the
($M_{V}$,$~(U-V)_{0}$) diagram have been performed under
assumptions of unimodal and bimodal mass distributions.

The statistical importance of the observed gap in the blue HB at
$\ell _{\mathrm{HB}}=0.42$ has been studied by simulations. We
conclude that it is a statistical fluctuation. \label{summary}

\section*{Acknowledgments}

We are grateful to the anonymous referee for his comments and
suggestions. \label {Acknowledgements}

\appendix

\section{Definitions of the probabilities used to test statistical
significance of the observed gap}

For the sake of completeness, we present definitions of the
probabilities $P_{1}$, $P_{2}$, and $P_{3}$ used in \S
\ref{sect33}. We restrict our attention only to the $\left( W,\ell
_{\mathrm{HB}}\right)$ relation.
\[
P_{1}=\int \left( 1-\int\limits_{0}^{W}m_{s}\left(
\overline{W}\right) d\overline{W}\right) \,m_{o}\left( W\right)
\,\,dW\,,
\]
\[
P_{2}=\int \int \frac{\left( 1-\int\limits_{0}^{W}f_{s}\left(
\overline{W},\ell _{\mathrm{HB}}\right) d\overline{W}\right)
}{m_{s}\left( \ell _{\mathrm{HB}}\right) }f_{o}\left( W,\ell
_{\mathrm{HB}}\right) \,\,dW\,d\ell _{\mathrm{HB}}\,,
\]
where $f_{o}\left( W,\ell _{\mathrm{HB}}\right) $ and $f_{s}\left(
W,\ell _{\mathrm{HB}}\right) $ are joint PDFs of the observed and
synthetic gaps, $m_{s}\left( W\right) $, $m_{s}\left( \ell
_{\mathrm{HB}}\right) $ and $m_{o}\left( W\right) $ are the
marginal PDFs of the $f_{s}\left( W,\ell _{\mathrm{HB}}\right) $
and $f_{o}\left( W,\ell _{\mathrm{HB}}\right) $. The integration
is performed over domain in $\left( W,\ell _{\mathrm{HB}}\right) $
plane with $f_{o}\left( W,\ell _{\mathrm{HB}}\right) >0$.
\[
P_{3}=\int \int \,f_{s}\left( W,\ell _{\mathrm{HB}}\right)
\,dW\,d\ell _{\mathrm{HB}}
\]
with a region of integration determined by the inequality
\[
f_{s}\left( W,\ell _{\mathrm{HB}}\right) <\int \int \,f_{s}\left(
W,\ell _{\mathrm{HB}}\right) f_{o}\left( W,\ell
_{\mathrm{HB}}\right) \,\,dW\,d\ell _{\mathrm{HB}}\,.
\]
\label{AppA}

\end{document}